**A site-controlled quantum dot system offering both high uniformity and spectral purity**

L.O. Mereni, V. Dimastrodonato, R.J. Young and E. Pelucchi

*Tyndall National Institute, University College Cork, Cork, Ireland*

In this letter we report on the optical properties of site-controlled InGaAs quantum dots with GaAs barriers grown in pyramidal recesses by metalorganic vapour phase epitaxy. The inhomogeneous broadening of excitonic emission from an ensemble of quantum dots is found to be unusually narrow, with a standard deviation of 1.19 meV, and the spectral purity of emission lines from individual dots is found to be very high (18-30 μeV), in contrast with other site-controlled dot systems.

Semiconductor quantum dots (QDs) are currently of great scientific interest finding application in a vast array of fields spanning fundamental physical studies and cavity quantum electrodynamics, to reach the latest quantum information technologies.[1] The ability to control, with nanometre precision, the location at which quantum dots are formed while maintaining the high optical quality which has, as far as now, always been a prerogative of self-assembled growth methods, is an open challenge for researchers. Under non-resonant optical excitation, the exciton line of single self-assembled QDs grown under the Stransky-Krastanov growth mode with molecular beam epitaxy (MBE) is usually broadened by the coupling of the QD excitons to phonons and to fluctuating charges in its environment. However, a small emission linewidth close to the limit imposed by the uncertainty principle can be observed sometimes at low temperature, provided high resolution set-ups and appropriate pumping conditions (e.g. pumping with a small excess energy) are used.[2,3] For site-controlled systems the spectral purity of the single dot excitonic emission is typically poorer than in self-organised systems and producing <50 μeV linewidths[4,5] in a reproducible way is challenging, though some progress has been made recently.[6] This is attributed to unintentional impurity incorporation that is associated with the pre-growth lithographic process needed to achieve the site-controlled nucleation. Moreover, metalorganic vapour phase epitaxy (MOVPE), which relies on growth precursors and has working pressures far from the ultra-high-vacuum conditions used for MBE, has failed, so far, to produce very narrow excitonic linewidths even when self-organised growth is employed.[7]

In this Letter we introduce a site-controlled quantum dot system grown by MOVPE in large (7.5 μm) inverted tetrahedral recesses on (111)B GaAs substrates. We present an evolution to the InGaAs pyramidal QD system, which raised considerable interest in the community particularly because of its high uniformity.[8] We show that it is possible to replace the $Al_yGa_{1-y}As$ barrier material with GaAs and that the growth process can then be performed at a relatively high temperature. This delivers improved epitaxial quality and, critically, preserves the delicate equilibrium between growth rate anisotropy and capillarity processes that leads to a self limiting profile and, finally, dot formation.[4] We present the results of optical studies of this system, in particular we show: the ability to tune the emission wavelength, high uniformity of the emission energy of the neutral exciton, and a neutral exciton emission of unprecedented spectral purity for single quantum dots grown by MOVPE.

The samples used in this study were grown by MOVPE on a (111)B GaAs substrate patterned with a matrix of tetrahedral recesses with a 7.5 μm pitch. These recesses, whose inner surfaces are oriented along the (111)A crystallographic directions, are obtained by standard lithography and wet chemical etching.[4] Growth conditions, temperature (730 ºC thermocouple), V/III ratio (in the range 600-800), and the material structure surrounding the quantum dot layer were nominally the same for all samples. Following a buffer layer of GaAs, $Al_yGa_{1-y}As$ was grown with a gradually varying composition from $Al_{0.30}Ga_{0.70}As$ to $Al_{0.75}Ga_{0.25}As$. Next a layer of $Al_{0.75}Ga_{0.25}As$ was grown to act as an etch stop to aid the post-growth process of substrate removal (back-etching)[4] which is employed to allow optical studies of the pyramids in the apex-up geometry. Finally the quantum dot and barrier layers are grown sandwiched between $Al_{0.55}Ga_{0.45}As$ cladding to provide carrier confinement. The width of the GaAs barrier beneath (above) the dot layer was 100 (70) nm. Four samples (A-D) were grown for this study each with a different composition of $In_xGa_{1-x}As$ in their respective 0.5 nm thick quantum dot layers; for samples A, B, C and D the compositions respectively were, x = 0.15, 0.25, 0.35 and 0.45. The sample structure following the etch-stop layer is illustrated in Fig.1, the vertical quantum wire[4] that forms through the centre of the structure is illustrated. While comprising a similar material system, this design is physically different to the one reported recently in Ref. 9. In the work by Gallo et al. the dimensions of the pyramidal template are >20 times smaller than in ours and, critically, on the order of the adatom diffusion length, thus the growth dynamics in their system and results they find are quite different to those presented here.

We characterized our structures optically by measuring micro-Photoluminescence (μPL) with non-resonant excitation at 658 nm. Samples were studied in two different geometries: "as-grown", with the apexes of the pyramids facing down, and, following substrate removal with a selective etch, "apex-up". The latter increases the proportion of the emitted light which is collected by our microscope objective by three orders of magnitude; the process also induces a red-shift in the quantum dots' emission of ~20nm.[4] The system resolution was ~10 μeV (2 pixels) at 800 nm and the sample were studied at <10K in a helium closed-cycle cryostat.

The average emission wavelength of the neutral exciton from a single dot for the four samples in the as-grown geometry is shown in Fig. 1(inset). As the emission wavelength for sample A coincides with emission from GaAs the value shown in the figure was extrapolated from a measurement in the apex-up geometry. As the concentration of Indium in the $In_xGa_{1-x}As$ alloy across the four samples was increased a clear shift in the emission wavelength from ~830 nm ($In_{0.15}Ga_{0.85}As$) to 882 nm ($In_{0.45}Ga_{0.85}As$) was observed. This is attributed to an increase in the depth of the potential well in which the excitons' wavefunction is confined. The wavelength shift of the emission is much smaller than would be expected naively in comparing the bandgaps of the $In_xGa_{1-x}As$ compositions of the four samples. This implies that a small fraction of the excitons' wavefunction resides in the dot; this will be the focus of theoretical studies of this system in the future.

The uniformity in the emission wavelength from sample B in the as-grown geometry is illustrated in Fig. 2 which plots the neutral exciton energy for *a large number* of individual pyramids. The average emission energy of the neutral exciton line, which was identified by mean of power dependencies and exchange splitting measurements, was found to be 1463.5 meV with a standard deviation of only 1.19 meV, the full width half maximum (FWHM) of the corresponding interpolated Gaussian fit is 2.8 meV which is slightly smaller than the results obtained in Refs 6 and 8.

Prior to substrate removal, in the as-grown geometry, the excitonic emission of a single QD is found to be relatively broad, with linewidths up to 90 μeV found on all the samples. In this geometry both the input- and output-coupling efficiencies are poor and a relatively large laser power density (>30 nW/μm$^2$) is required for non-resonant excitation. Following substrate removal the linewidth of the emission is found to be significantly reduced. Fig. 3 shows two representative μPL spectra of neutral exciton (X) and biexciton (2X) emission collected from individual quantum dots pumped with low excitation power densities. Linewidths of the neutral exciton lines were measured by fitting the spectra to Lorentzian lineshapes and found to be 18.0 μeV for (a) and 19.4 μeV for (b).These results are comparable with those of best MBE growths of self-organised quantum dots and are significantly better than the previous best results reported for site-controlled QDs.[4,5,8] As the pump power is increased other lines, associated to multiexcitonic complexes, appear in the spectrum (not shown).

In conclusion we have introduced a site-controlled quantum dot system, composed of an $In_xGa_{1-x}As$ dot embedded in GaAs, demonstrating both high uniformity and spectral purity. Varying the composition of the $In_xGa_{1-x}As$ quantum dot allowed the emission energy of the neutral exciton to be tuned which could be of great value to experiment and promises development of a long-wavelength QD system for quantum information applications.

This research was enabled by the Irish Higher Education Authority Program for Research in Third Level Institutions (2007-2011) via the INSPIRE programme, and by Science Foundation Ireland under grants 05/IN.1/I25 and 05/IN.1/I25/EC07. We are grateful to K. Thomas for his support with the MOVPE system.

**Figure Captions**

Fig. 1: (colour online) An illustration of the sample layer structure following the etch-stop layer used to facilitate substrate removal. Label 1 – $Al_{0.55}Ga_{0.45}As$ layer, 2 – GaAs barriers, 3 – $In_xGa_{1-x}As$ dot layer, 4 – Vertical quantum wire. Inset: Emission energy tuning as a function of indium percentage in the quantum dot composition.

Fig. 2: The emission energy of the neutral exciton emission measured for a large number of pyramids from sample B. The solid line marks the average value, the two dashed lines mark the FWHM of the Gaussian fit.

Fig. 3: μ-Photoluminescence spectra taken from two different individual quantum dots. Emission from the neutral exciton (X) and biexciton (2X) states is labelled. The width of the neutral exciton peaks were extracted by fitting a lorentzian peak. Integration times and excitation intensities are different for the two spectra.

**Figure 1**

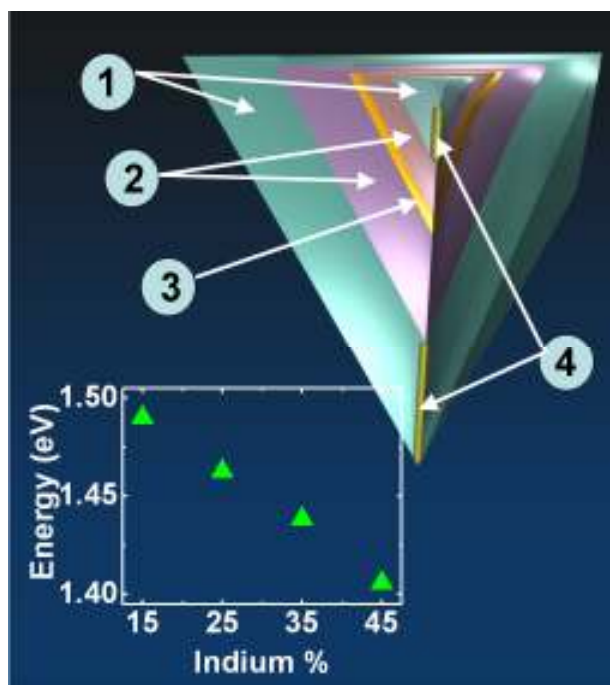

**Figure 2**

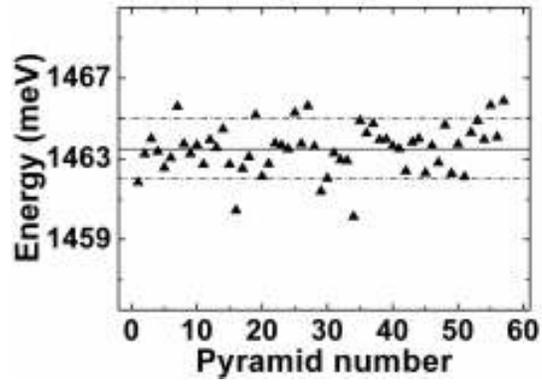

**Figure 3**

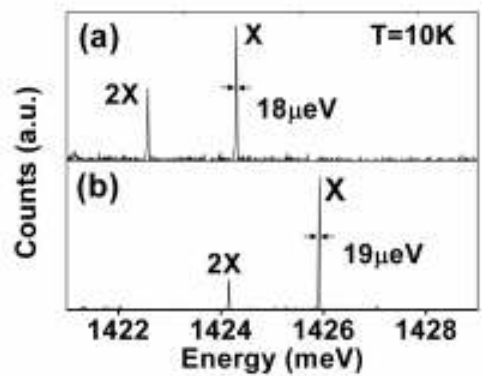